\documentclass[twocolumn,tighten]{aastex631}
\usepackage{graphicx}

\usepackage{xcolor}
\PassOptionsToPackage{mathlines}{lineno}
\usepackage{amsmath}

\shorttitle{Evidence for Millihertz Oscillations in the bright atoll source GX 3+1}
\shortauthors{Malu et al.}

\graphicspath{{./}{figures/}}

\begin{document}
\makeatletter
\def\NAT@sort{1}
\makeatother
\title{Evidence for millihertz quasi-periodic oscillations in the bright atoll source GX 3+1}

\author[0000-0003-0440-7978]{Malu Sudha}
\affiliation{Department of Physics \& Astronomy, Wayne State University, 666 West Hancock Street, Detroit, MI 48201, USA}
\author[0000-0002-8961-939X]{Renee M.\ Ludlam}
\affiliation{Department of Physics \& Astronomy, Wayne State University, 666 West Hancock Street, Detroit, MI 48201, USA}
\author[0000-0002-3422-0074]{Diego Altamirano}
\affiliation{School of Physics and Astronomy, University of Southampton, Southampton, Hampshire SO17 1BJ, UK}
\author[0000-0002-8294-9281]{Edward M.\ Cackett}
\affiliation{Department of Physics \& Astronomy, Wayne State University, 666 West Hancock Street, Detroit, MI 48201, USA}
\author[0000-0002-0940-6563]{Mason Ng}
\affiliation{Department of Physics, McGill University, 3600 rue University, Montréal, QC H3A 2T8, Canada}

\affiliation{Trottier Space Institute, McGill University, 3550 rue University, Montréal, QC H3A 2A7, Canada}



\begin{abstract}

We report evidence for millihertz (mHz) QPOs in the bright atoll neutron star low-mass X-ray binary (NS LMXB) source GX 3+1 using NICER in the 0.5--10 keV energy band. Across 7 observational datasets obtained over 6 days, we made 8 candidate mHz QPO detections with local significance above 95\%, one of which remains above 95\% global significance after the trial correction. These mHz QPOs were detected in the 7--15 mHz frequency range and had fractional rms amplitudes ranging from 0.51--1.41\%. Our studies indicate no association between the hardness intensity diagram location of the candidate QPOs and their rms amplitudes. There appears to be a monotonous increase in rms with energy, except in some observations where there is a pivot around 3--4 keV or at $\sim$ 5 keV. Previous studies of mHz QPOs in other sources in literature indicate a pivot around 3 keV in the rms-energy relation, but in our study some tentative detections suggest a pivot at around $\sim$5 keV, though the large uncertainties in most cases prevent a robust statistical claim. Although properties such as the frequency range of detection and fractional rms amplitudes of the mHz QPOs in our study are well in agreement with that in literature for other NS LMXBs, the luminosity at which these candidate QPOs occur are higher than that of other sources. The rms-energy relation of the candidate mHz QPOs and the luminosity at which they occur challenges some of the existing considerations of mHz QPO origin. 
\end{abstract}


\section{Introduction} \label{sec:intro}

A binary system hosting a neutron star (NS) accreting material from a low-mass companion star ($<$ 1 M$_{\odot}$) with the accreted material forming an accretion disk via Roche-lobe overflow is termed as a NS low-mass X-Ray binary (LMXB). Primarily, they are classified into two broad categories called `Z' and `atoll' sources. This classification is based on the track they trace out in the hardness intensity diagram (HID) or the color-color diagram (CCD) and the correlated timing features (see \citealt{hasinger1989}). While Z sources exhibit high X-ray luminosity ranging from $L_x\sim$ 0.5–1~$L_{\rm Edd}$, atoll sources have relatively lower X-ray luminosity with $L_x\sim$ 0.001–0.5~$L_{\rm Edd}$. Z sources trace out three separate branches in their HID/CCD which are referred to as the horizontal branch (HB), normal branch (NB), and flaring branch (FB) \citep{hasinger1989}. Atoll sources exhibit three main branches termed as the extreme island state (hard state), island state (intermediate state) and the banana branch (soft state)\citep{hasinger1989,van2006}. The source XTE J1701–462 switched between Z and atoll subclasses at high luminosities which made the distinction between the subclasses less clear \citep{lin2009,homan2010}. This indicated that mass accretion rate could be the governing factor that distinguishes between these classes (see also \citealt{altamirano2010,oosterbroek1995,wijnands1999, ng2024}). These systems also exhibit Type-I X-ray bursts that mark a fast rise and a relatively slower decay in the X-ray flux due to thermonuclear runaways occurring in the freshly accreted hydrogen and/or helium on the NS surface (e.g., \citealt{lewin1993,strohmayer2006,galloway2008}).

Both Z and atoll sources show distinct timing characteristics associated with each spectral state. In Z sources, the HB, NB, and FB are known to exhibit characteristic quasi periodic oscillations (QPOs) called HB oscillations (HBOs), NB oscillations (NBOs), and FB oscillations (FBOs). HBOs occur in the 15-60 Hz frequency range, while NBOs and FBOs occur in the 5-8 Hz and 10-25 Hz frequency ranges \citep{van2006, hasinger1989}. These systems also exhibit kHz QPOs \citep{van1996}.
Atoll sources also exhibit a wide range of QPO frequencies such as the millihertz (mHz) QPOs \citep{revnivtsev2001,heger2007}, $\sim$ 1 Hz class of QPOs \citep{homan2015}, hectohertz QPOs (e.g.  \citealt{altamirano2008}) and kHz QPOs \citep{van1996}. Apart from these they also exhibit peaked noise features and both low and high frequency broadband noise features \citep{van2006}. Low-frequency QPOs are often associated with inner accretion flow and disk instabilities, while QPOs in the mHz frequency range are associated with NS surface phenomenon like the marginally stable nuclear burning process of Helium (MSNB; \citealt{revnivtsev2001, heger2007}). 

There are multiple spectral models that have been employed to explain the observed emission of NS LMXBs. While \cite{white1988} used a single temperature blackbody emission from the NS or boundary layer (BL) and Comptonized emission from the corona in the inner disk region (`Western model') to explain it, \cite{mitsuda1984} used a multi-temperature disk blackbody emission and a Comptonized corona for describing the observed emission (`Eastern model'). Using the `hybrid' spectral model by \citealt{lin2007}, the continuum spectrum of the hard state in atoll sources could be described using a single-temperature blackbody for the boundary layer and a power-law component for the Comptonization component. The continuum spectrum of the soft state on the other hand could be described using a multi-temperature blackbody for the disk, a single-temperature blackbody for the BL and a power-law for the Comptonization component. Additionally, when the hard X-ray photons from the BL or the coronal region illuminates the disk, it leads to a reprocessed emission from the disk which is termed as the 'reflection' spectrum. A predominant feature of the reflection spectrum is the Fe K$\alpha$ line which can reveal critical information about the accretion flow \citep{fabian1989}. Modeling the reflection spectrum, especially the shape of the Fe line profile can help us determine the radius of inner accretion disk region based on the understanding that the Doppler and relativistic effects are stronger closer to the NS (see \citealt{ludlam2024} for a review). 

GX 3+1  is a known Type-I X-ray burster. Considering that these bursts are Eddington-limited, the source distance has been estimated to have an upper limit of $\sim$ 6.5 kpc \citep{galloway2008}. This is also consistent with the photospheric radius-expansion (PRE) based estimate \citep{kuulkers2000}. This source has shown signatures of a Fe K$\alpha$ line \citep{oosterbroek2001,piraino2012,ludlam2019} and  spectral modeling indicate an inclination angle of 28$^\circ$--44$^\circ$ \citep{fabian1989,ludlam2019,pintore2015}. This source has always been found in the soft banana branch in literature. \cite{lewin1987} detected a $\sim$ 8 Hz QPO and a low-frequency noise feature below 10 Hz in the EXOSAT observations of this source. This QPO having a rms percentage of $\sim$ 3 \% was interpreted to be similar in nature to NBO feature in Cygnus X-2, owing to its intensity-independence and the spectral hardness at which it occurs \citep{jia2023,sudha2025,zhang2025}. \cite{seifina2012} described the PDS of this source in the upper banana branch to be dominated by a very low-frequency noise with a break around 20 Hz and the lower banana branch to be dominated by a high-frequency noise in the 1-100 Hz range, along with a very low frequency noise below $\sim$ 1 Hz. 

Here we present the spectro-temporal studies of the persistently accreting bright atoll source GX 3+1 using the Neutron Star Interior Composition Explorer Mission (NICER) in the 0.5--10 keV energy band. A detailed spectro-temporal analysis of the source has previously never been performed in the extreme soft X-ray energies. NICER with its high temporal and energy resolutions can provide us with an unprecedented view of the spectro-temporal behavior of this source. In this paper, we discuss the results from the spectro-temporal studies of GX 3+1 using NICER data.

\section{Observation and Data Reduction}\label{sec:obs}

Seven different NICER observations from 2025-05-16 to 2025-05-22 were used for our study. The observation details are as given in Table \ref{tab1}. 
\begin{table}[!t]
    \centering
        \caption{NICER Observation Details of GX 3+1.}
    \begin{tabular}{cccccc}
        \hline
       Obs \# & Obs.\ ID & Obs.\ Date &  Exp.\ (ks) \\
       \hline
        1 & 8611010101 & 2025-05-16 19:54:20.00 & $\sim$ 0.5 \\
        2 & 8611010102 & 2025-05-17 00:46:00.00 & $\sim$ 4.7 \\
        3  & 8611010103 & 2025-05-18 03:07:40.00	 &$\sim$ 3.4 \\
        4  & 8611010104 & 2025-05-19 00:23:21.00& $\sim$ 4.7 \\
       5  & 8611010105 & 2025-05-19 23:51:39.00 &$\sim$ 4.9 \\
       6 & 8611010106 & 2025-05-21 01:56:42.00 &$\sim$ 3.7 \\
        7  & 8611010107 & 2025-05-22 01:12:07.08&$\sim$ 0.4 \\
        \hline
    \end{tabular}

    \label{tab1}
\end{table}
There are two additional archival datasets available for this source, but after the standard cleaning and filtering procedure (mentioned below) each spans $\sim$ 200 s. Therefore, owing to the very short duration of these observations, these were not used for this study. NICER data was reduced following the standard procedure using the `nicerl2' routine. Since these observations were performed after NICER experienced the optical light leak\footnote{https://heasarc.gsfc.nasa.gov/docs/nicer/analysis$\_$threads/light-leak-overview/}, we used only the orbit night data for the remaining analysis. Filtering criteria of KP $<$ 5 and COR\_SAX $>$ 4 were applied to reduce the particle background at lower energies (e.g., \citealt{ludlam2022}). Barycentric correction was performed using the `barycorr' tool in the ICRS reference frame using the RA/DEC coordinates of RA 266.9846 and DEC -26.5631. This resulted in a barycentered cleaned event file. The `nicerl3-spect' routine and the SCORPEON model were used to obtain the spectra and estimate the background in our spectral modeling\footnote{\href{https://heasarc.gsfc.nasa.gov/docs/nicer/analysis_threads/scorpeon-overview}{scorpeon-overview}} respectively. 
The spectra were not binned leaving the native bin size of 10 eV across the entire spectral range. Background lightcurves estimated from the SCORPEON model showed a count rate $<$ 1~count~ s$^{-1}$ compared to the $\sim$ 1100-1500~counts~s$^{-1}$ source count rate, which led to the background not being subtracted, as it is negligible compared to the source count rate (e.g., \citealt{zhang2021}). Timing analysis was performed using the Stingray\footnote{https://docs.stingray.science/en/stable/} software suite \citep{huppenkothen2019,huppenkothen2019a,bachetti2021}. Each individual good time interval (GTI) of the NICER lightcurves were analyzed separately in a search for QPO or noise features.

\begin{figure*}
\centering
\includegraphics[width=\textwidth, angle=0]{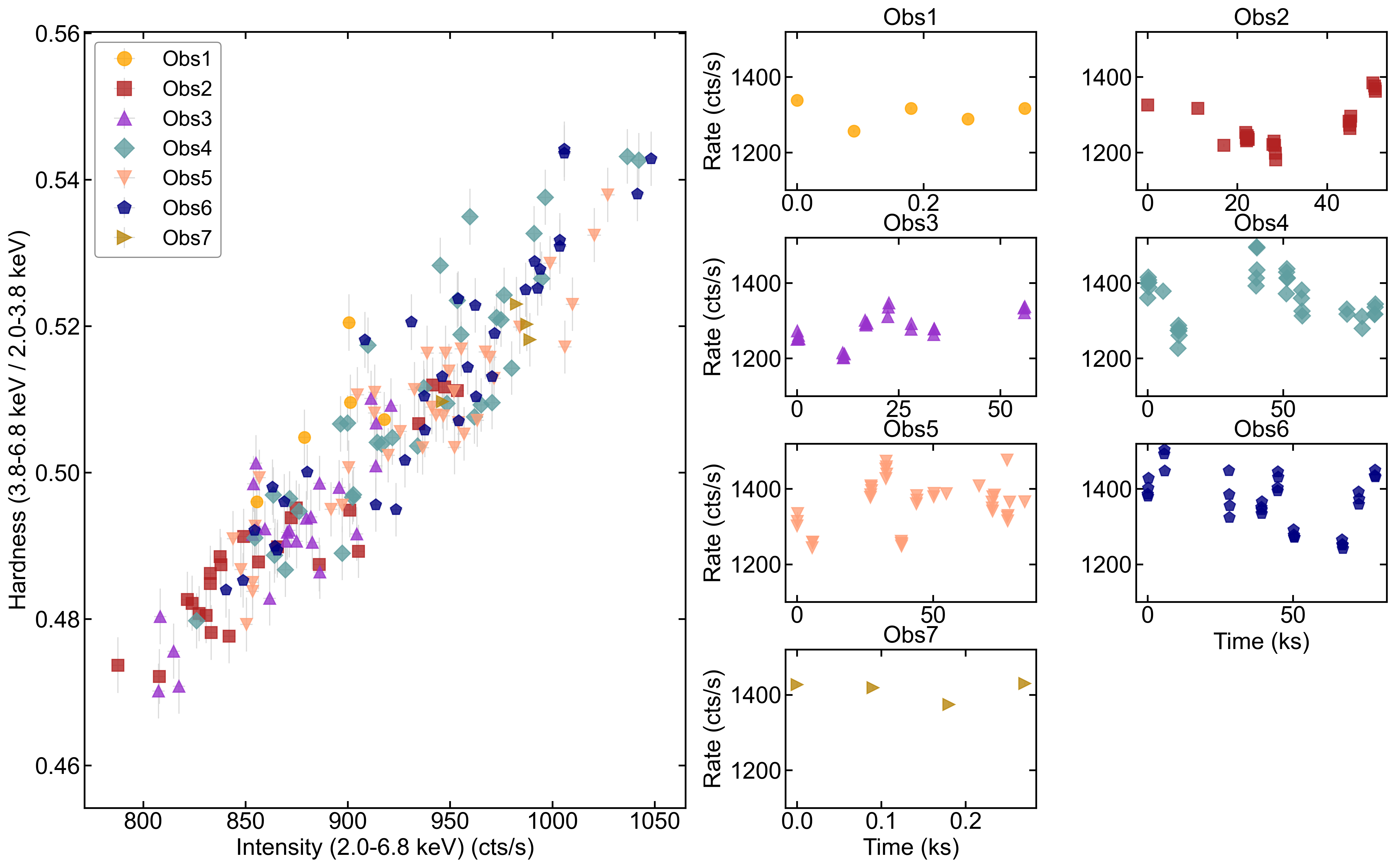}
\caption{Left: Hardness intensity diagram of GX 3+1 obtained using NICER Obs1-7. Hardness ratio is the ratio of X-ray photon countrate between the 3.8--6.8 keV and 2--3.8 keV energy bands and intensity is the total countrate in the 2--6.8 keV energy band. HID is obtained with data binned to 90 s. Right: Lightcurves in the 0.5--10 keV energy band binned to 90 s using the same styled markers as used in the HID.}
\label{fig:hid}
\end{figure*}

From here on, we refer to the NICER observations from May 16-22 as Obs1-Obs7 (Table \ref{tab1}). We obtained the HID using NICER lightcurves binned to 120 s as shown in Figure \ref{fig:hid}, where hardness ratio is the ratio of X-ray photon countrate between the 3.8--6.8 keV and 2--3.8 keV energy bands and intensity is the total countrate in the 2-6.8 keV energy band. Figure \ref{fig:hid} also shows the observation-wise lightcurves binned to 90 s in the 0.5--10 keV energy band. The mean count rate of lightcurves vary from $\sim$ 1270--1413 cts/s. The source is considered to be in the soft banana state based on the obtained HID which shows characteristic positive correlation between hardness and intensity as seen in atoll sources (e.g. \citealt{hasinger1989}, \citealt{asai1993}, \citealt{neal2023}). This is further supported by previous studies in literature that found the source to exist only in the soft banana state \citep{asai1993,mondal2019,neal2023}.

\section{Results}\label{sec:results}
\subsection{Timing Analysis}
 
Power density Spectra (PDS) in the 0.5--10 keV energy band were created using STINGRAY's AveragedPowerspectrum routine for each observation by averaging PDSs of lightcurve segments with 32 s length and a time bin size of 1/256 s. This resulted in a Nyquist frequency of 128 Hz and frequency resolution of 0.03125~ Hz (i.e 1/32 s$^{-1}$). The averaged PDS were fairly featureless. 

We also explored timing features in the millihertz frequency domain as atoll NS LMXBs are known to exhibit millihertz QPOs in the 2--20 keV energy band. We generated Lomb-Scargle periodograms \citep{lomb1976,scargle1982} using 1 s time resolution lightcurves for all GTIs longer than 400 s. The LombScarglePowerspectrum routine from STINGRAY was utilized for this purpose and Leahy normalization was used for the power spectra. We searched a frequency range of 5-15 mHz for possible mHz QPOs. 

We detected candidate mHz QPOs in several GTI segments. We further tested these tentative detections by estimating their significance using the procedure followed by \cite{vaughan2005} and also using Monte-Carlo simulations. 

Following the Vaughan method, we first modeled the logarithmically rebinned periodogram continuum using a power-law model (for the low frequency red noise) and a constant $=$ 2 for the Poisson noise of the Leahy normalized PDS, after ignoring the frequency range of interest (5-15 mHz). To prevent the continuum from being biased by the noise at higher frequencies, a median absolute deviation (MAD) filter was applied where points more than 3 $\times$ MAD from the median power were excluded. Then we determined the 95 \% significance threshold estimated by considering the fact that the ratio 2$\times$ (signal power/noise power) follows a $\chi$$^{2}$ distribution with 2 degrees of freedom and by taking into account the number of independent trials in the periodogram per GTI segment and by also considering the number of independent segments analyzed. 

We also estimated significance of the candidate mHz QPO detections using Monte-Carlo simulations, where we employed the \cite{timmer1995} method using DELCgen package \citep{timmer1995,connolly2015}. For each GTI, 5000 lightcurves were simulated from the best fit continuum model and then a Lomb-Scargle periodogram were generated for each simulated lightcurve. Significance was estimated from the distribution obtained from this simulation. From the Lomb–Scargle periodogram, we extracted the maximum power within the QPO search band. QPOs were considered significant only if they had $>$ 95 \% significance. 

Unfortunately, we did not detect any Type-I X-ray bursts in our datasets, which could be due to limitations of the data with several short GTI segments. This prevents us from searching for any association of the candidate mHz QPOs with any burst occurrence. Furthermore, data statistics limits us from performing any search for frequency drift or QPO phase resolved studies.

This is the first detection of mHz QPOs in GX 3+1. We have identified 8 candidate mHz QPOs based on our analysis with frequencies ranging from 7-15 mHz (see Table \ref{tab:qpo_properties} for details). This is based on the local significance estimate. The local significance accounts for the number of independent frequency trials within a single GTI segment. The number of independent frequency trials per segment N$_{local}$ $\sim$ 2T$_{GTI}$~$\Delta$f$_{range}$ where $\Delta$f$_{range}$=10mHz is the search bandwidth (5–15 mHz) and T$_{GTI}$ is the GTI length. Across our 16 segments, GTI lengths range from 400–585 s (minimum length of 400 s), yielding 8 - 11 independent frequency trials per segment. 
The global significance additionally corrects for the total number of independent trials across all segments searched. We have N$_{global}$ = 145 independent trials in total. Applying this global trial correction, we obtain 7 tentative detections at $\geq$ 68\%, 4 at $\geq$90\%, and 1 at $\geq$95\% global significance. 

The GTI lengths correspond to frequency resolutions of $\Delta$f=1/T=1.71–2.50 mHz and cycle counts of N=4–7. With only 4–7 cycles per segment, the frequency of each detection is uncertain at the level of $\Delta$ f and we note that this limits robust QPO identification. Furthermore, with so few cycles, a red-noise fluctuation could in principle mimic a QPO peak. To guard against this, all detections were required to simultaneously satisfy two independent significance criteria as mentioned above for red-noise significance against a fitted power-law continuum, and a Monte Carlo test at $\geq$ 2$\sigma$, strongly disfavoring red-noise fluctuations. To further assess robustness, we tested the stability of each tentative QPO detection by recomputing the periodogram on segments merging consecutive GTIs within the same ObsID. Among the eight tentative detections, for two detections, no usable adjacent GTIs exist, precluding a stability test. For the remaining six, the signal is recovered within $\leq$ $\Delta f$, except one detection which showed a frequency drift of $\sim$2.9 mHz between orbits. 

The Lomb–Scargle periodogram was adopted as the primary analysis tool owing to its finer frequency sampling grid, which allows for improved peak localization. By utilizing an oversampling factor of 3, it yields a significantly higher density of frequency bins compared to the native Fourier grid, which provides only 4–5 coarse frequency bins across the entire 5–15 mHz search band. We utilized the oversampling grid strictly for precise peak localization. For our significance thresholds, the analytical estimates were based on the number of independent Fourier frequencies. Meanwhile, the Monte Carlo significance estimates were evaluated using the same oversampled frequency grid as the observational data, thereby naturally accounting for the correlations between adjacent oversampled frequencies. We verified that all eight tentative detections are nonetheless reproduced by a standard Fourier PDS obtained using STINGRAY. The Fourier peak frequencies agree with the Lomb–Scargle values in all cases and the Vaughan significances are consistent between the two methods, confirming that the tentative detections are not an artifact of the choice of periodogram. Furthermore, we note that the GTI structure itself introduces a window function whose properties are particularly relevant at the very low frequencies considered here. Since each segment is a single contiguous GTI of duration T=400--585 s, the frequency resolution limit of 1.71--2.50 mHz, which lies below the search band minimum of 5 mHz, ensuring that the entire search band is fully resolved by each individual segment. Since we analyze each GTI independently rather than evaluating a globally concatenated time series, data gaps do not enter the spectral window function of any individual periodogram. The only window-function effect relevant to our analysis is therefore the finite-segment broadening of any true signal. Across our 5--15 mHz search band, this corresponds to a fractional frequency uncertainty of $\Delta f / f \approx 11-50\%$.

\begin{figure}
\centering
\includegraphics[width=0.47\textwidth, angle=0]{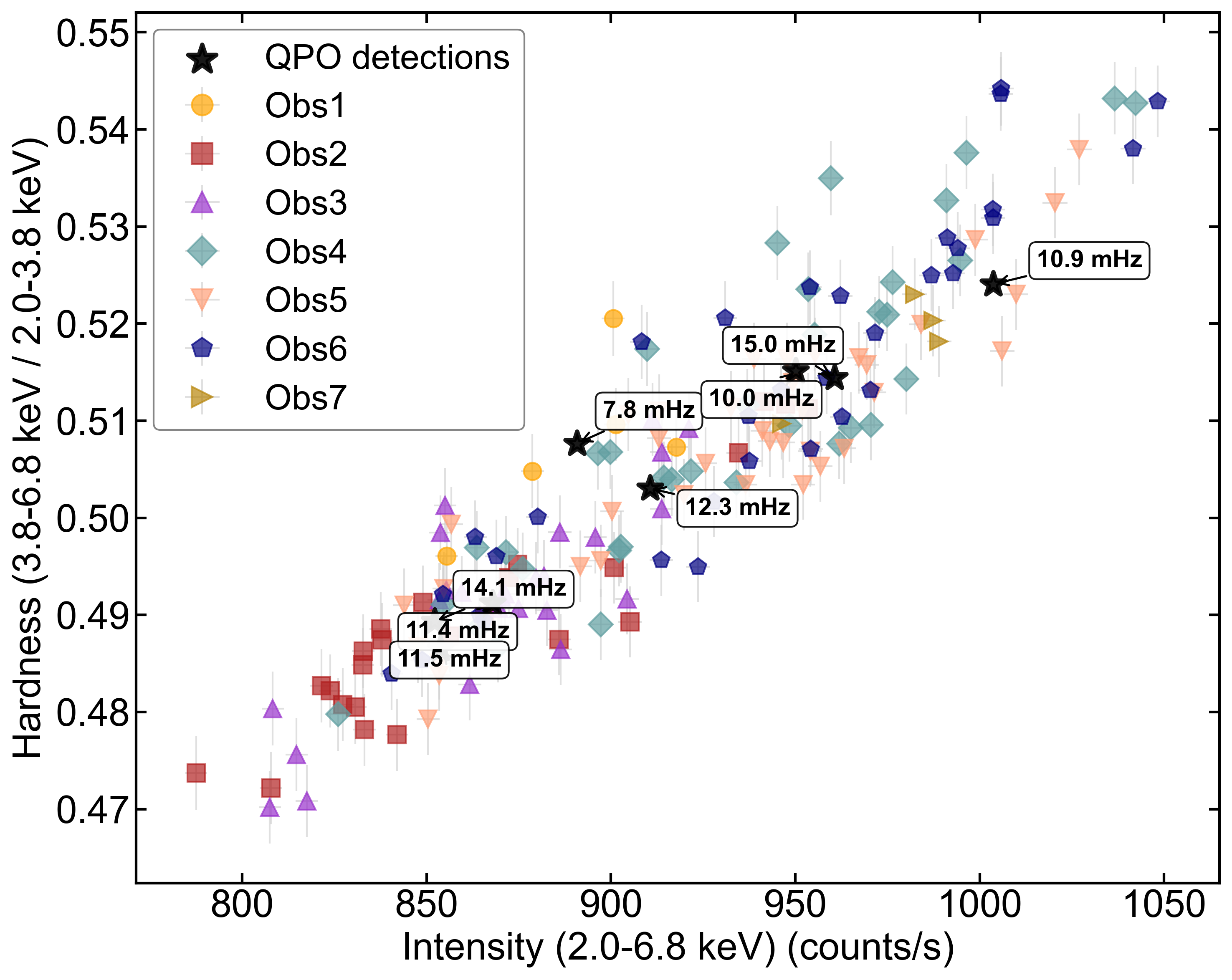}
\caption{HID of GX 3+1 (grey markers) obtained using NICER Obs1-7 overplotted with the location of the candidate mHz QPOs shown using the black `*' marker.}
\label{fig:hidqpo}
\end{figure}

We traced the location of these candidate mHz QPOs in the HID as shown in Figure \ref{fig:hidqpo}. As seen in the figure, these tentative mHz QPO detections are not localized to any specific region on the observed HID. But we refrain from drawing any conclusions based on this, since we cannot determine the spread of the HID.
\begin{table}[!ht]
\centering
\footnotesize
\caption{Candidate mHz QPOs in GX 3+1. RMS is estimated by fitting the folded lightcurves using a sinusoidal function and a constant, with rms=A/$\sqrt{2\times (C-B)}$, where A is the amplitude of oscillation of the fundamental frequency, C is the constant level, and B is background count rate. Errors reported are 1$\sigma$ error values. For the fundamental frequencies, \(\pm \Delta \nu\) denotes the symmetric half-width uncertainty imposed by the finite segment length (\(\Delta \nu = 1/2T\)). $\chi^2$/dof gives the fit statistics of the sinusoidal model fit to the phased profile. The asterisk on the RMS estimate of Obs1 indicates that the fit has a high $\chi^2$/dof owing to its asymmetric profile leading to an unreliable estimation of the fractional RMS amplitude.}

\label{tab:qpo_properties}
\begin{tabular}{lclcc}
\toprule
Observation  & Frequencies (mHz) & RMS (\%) & $\chi^2$/dof \\ &  $
\nu$ \(\pm \Delta \nu\)& (0.5--10 keV) \\
\hline
Obs1 &  ${7.8 \pm 1.0}$ & $1.41 \pm 0.12^{*}$ &2.71 \\
\hline
Obs2 &  ${11.4 \pm 1.0}$ & $0.54 \pm 0.12$ &1.26 \\
\hline  
Obs4 &  ${12.3 \pm 1.0}$ & $0.79 \pm 0.13$ &1.28 \\
\hline
{Obs5} &  ${10.9 \pm 1.1}$ & $0.52 \pm 0.12$ &1.33\\
 &  ${11.5 \pm 1.2}$& $0.67 \pm 0.14$ &1.22\\
 &   ${15.0 \pm 1.1}$  & $0.58 \pm 0.13$&1.12 \\
\hline
{Obs6} &  ${10.0 \pm 1.3}$ & $0.61 \pm 0.14$&1.21 \\
 &  ${14.1 \pm 1.0}$ & $0.51 \pm 0.13$ &1.06\\
\hline\hline
\end{tabular}
\end{table}

\begin{figure*}[!t]
\includegraphics[width=\textwidth]{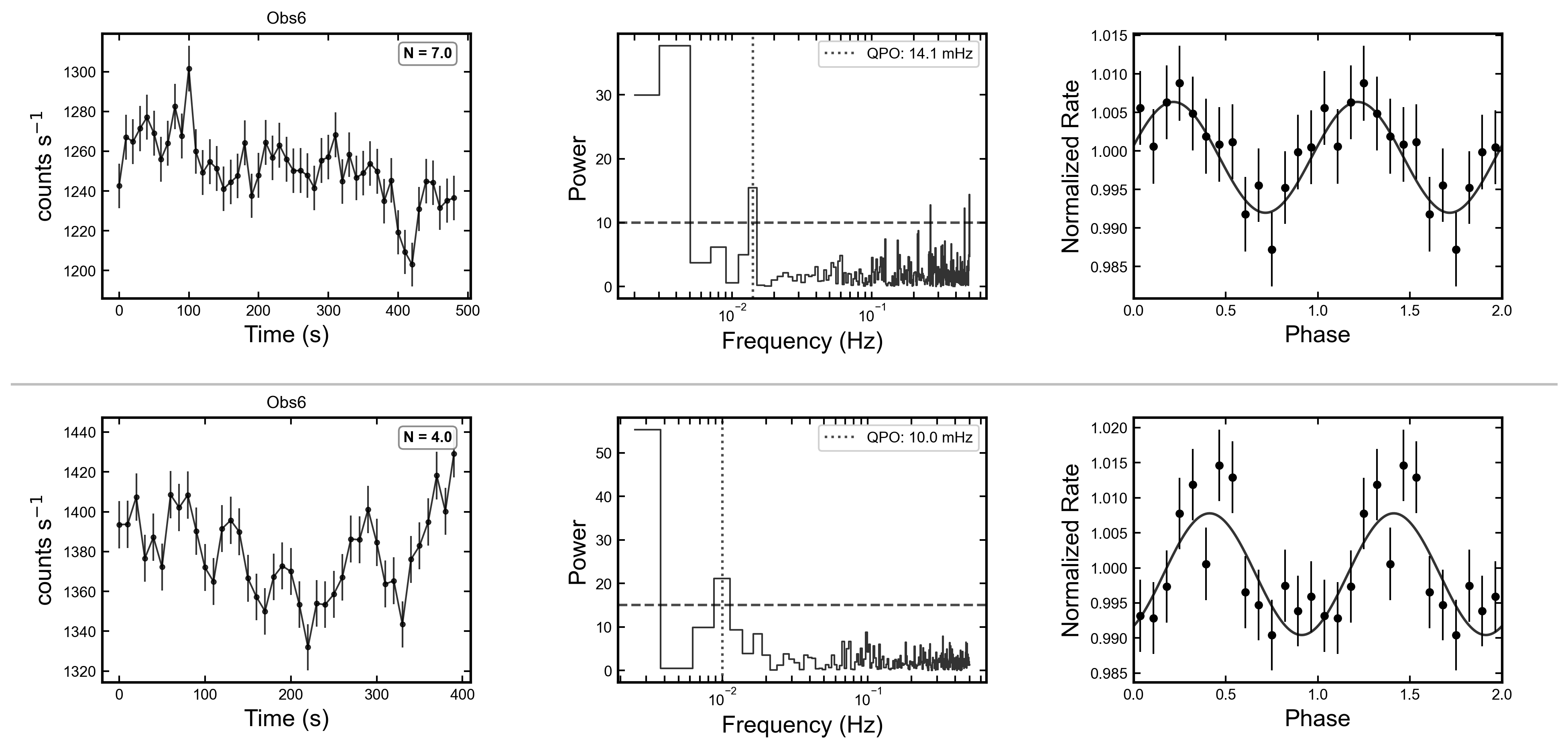}
\caption{The lightcurve segments that have tentative mHz QPO detections (left) with the number of QPO cycles in each segment (N), identified candidate mHz QPOs (center panel) and their corresponding folded profiles at the fundamental frequencies (2 cycles) with the solid black curve showing the sinusoidal fit to the profile (right panel) in the 0.5--10 keV energy band. Overplotted on the power spectra is the 95 \% significance threshold (black dashed line) from the Monte-Carlo simulations. The black vertical dotted line marks the central frequency of the candidate mHz QPO.}
\label{fig:qpodets}
\end{figure*}

We obtained the mHz QPO profiles by folding the lightcurve segments that exhibited these oscillations using the fundamental frequencies of the respective QPOs (Figure \ref{fig:qpodets}). This figure shows all the 8 tentative detections obtained upon considering a local significance $>$ 95 \% (left panel) along with the number of QPO cycles (N) in each segment. The right panel shows the profiles obtained folding the lightcurve segments with the obtained QPO frequency. We note that some of the QPO profiles are slightly asymmetric, possibly owing to the contributions from harmonic components \citep{heger2007}.

\begin{figure*}[!t]
\centering
\includegraphics[width=\textwidth]{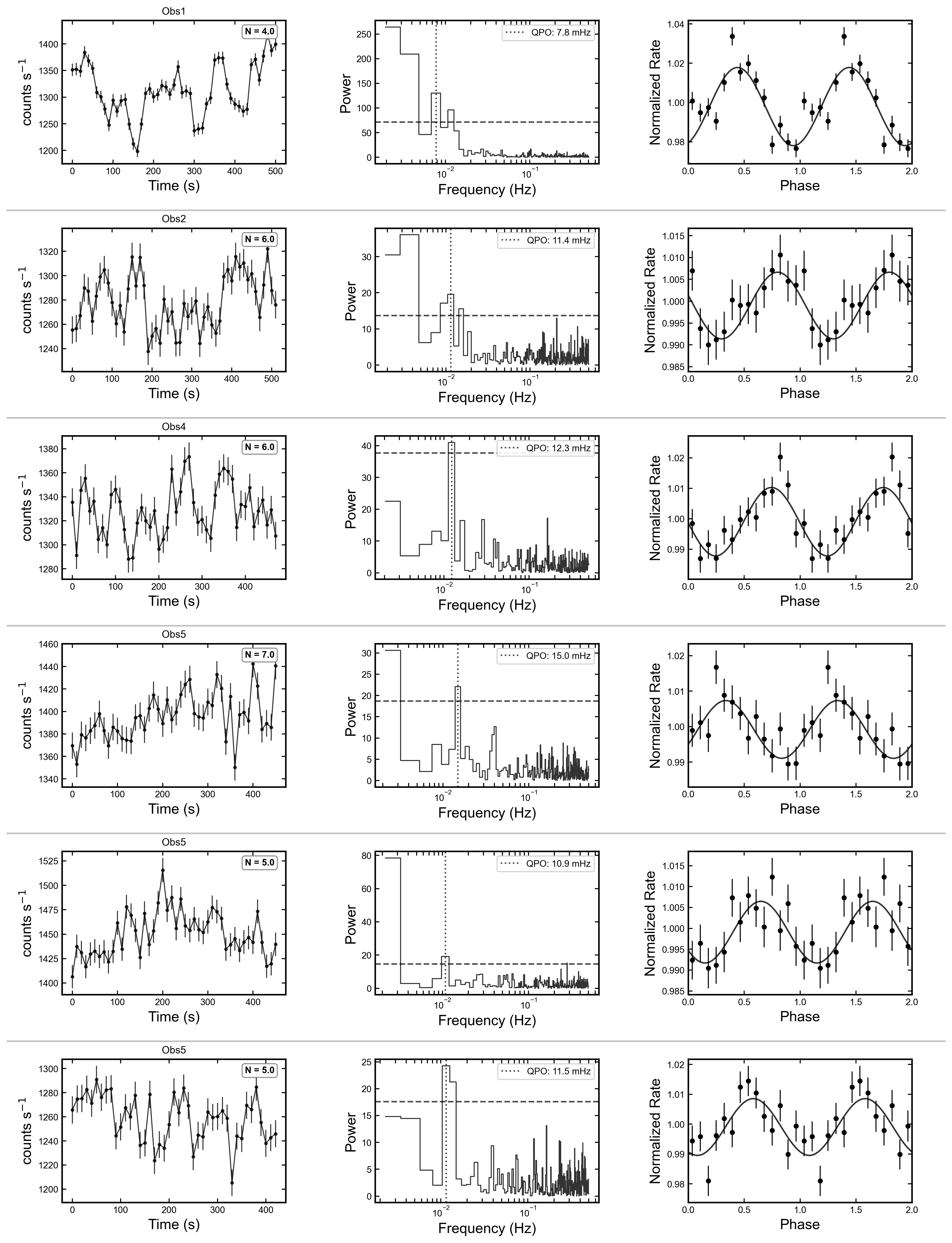}
\caption*{}
\end{figure*}

These folded lightcurves were then fitted using a model comprising a sinusoidal function and a constant. Fractional rms amplitudes of the candidate mHz QPOs, rms=A/$\sqrt{2\times (C-B)}$. Here A is the amplitude of oscillation of the fundamental frequency, C is the constant level, and B is background count rate which is approximated to 0 in this case owing to the negligible $<$ 1 cts/s background of the NICER lightcurves. The 1$\sigma$ errors on RMS are obtained by error-propagation of the fit covariance matrix. Obs1 profile has a very high $\chi^2$/dof owing to the asymmetric profile leading to a poor estimation of its fractional RMS amplitude, but it is still presented here for the sake of completeness. 

To determine whether the oscillation disappears entirely or merely drops below the detection threshold in non-detection segments, we computed 90\% confidence upper limits on the fractional rms amplitude in all non-detection segments satisfying T$\geq$ 400 s. The lightcurve of each non-detection segment was folded at the mean QPO frequency detected within the same ObsID, and the same sinusoidal model used for detection rms measurements was fitted. The 90\% upper limit on the oscillation amplitude estimated from this fit was used to obtain the rms upper limit. We note that most of the GTI segments within each observation are shorter than 400 s with many lasting only a few tens of seconds, therefore preventing any meaningful upper limits to be placed on these segments. For the 7 non-detection segments where upper limits could be estimated, we obtain rms upper limits of 0.5–1.1\% which is consistent with the rms amplitudes measured in the detection segments. This indicates that the oscillation amplitude in non-detection segments is comparable to that in detection segments and therefore intermittency could be likely attributed to a drop below detection threshold.

\begin{figure}
\centering
\includegraphics[width=0.48\textwidth, angle=0]{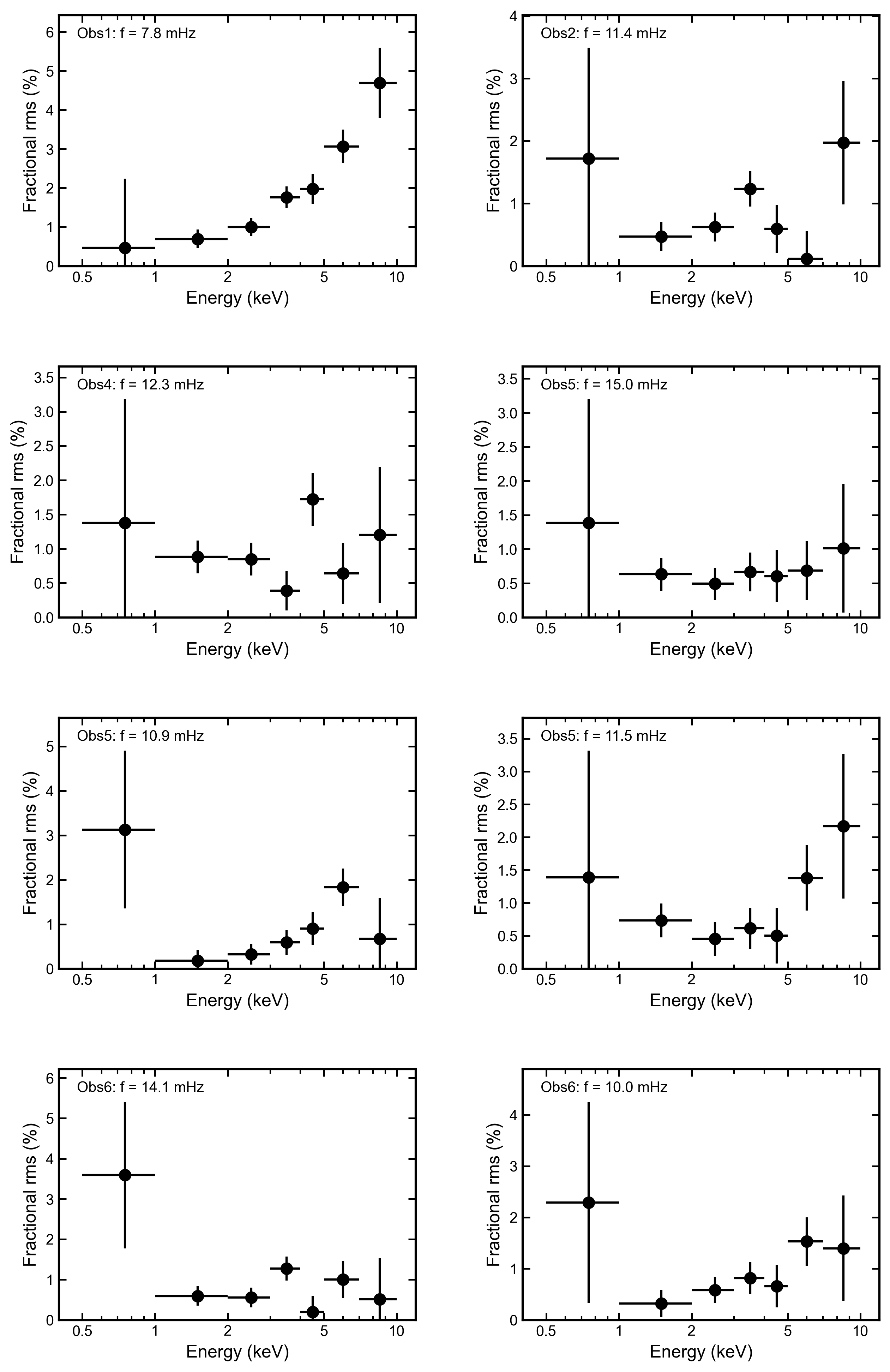}
\caption{Fractional rms of the candidate  mHz QPOs in different energy bands with the respective QPO frequencies mentioned on each panel.}
\label{fig:rmsvsenergy}
\end{figure}

We then looked for any specific trends in the rms percentage values along the HID and found no correlation with intensity or hardness. Further, we obtained the rms-energy dependence by obtaining fractional rms amplitude in different energy bands in the same manner as described above. Figure~\ref{fig:rmsvsenergy} shows the rms-energy curve for each detection shown in Table \ref{tab:qpo_properties}. We notice a monotonous increase in rms with respect to energy for the 7.8 mHz, 10 mHz, and 11.5 mHz QPOs above 1 keV. For the 10.9 mHz QPO rms increases with energy but only till $\sim$ 5 keV after which rms is constant within uncertainties. For the 11.4 mHz QPO, this pivot is around 3--4\,keV. For the 15 mHz and 14.1 mHz QPOs, the uncertainties are too large to confirm any specific trend. The increasing trend seen in the rms-energy curve of the 7.8~mHz QPO is unreliable owing to the high $\chi^2$/dof as mentioned above, but it is shown here for the sake of completeness.

\subsection{Spectral analysis}   \label{sec:spect}
To understand the associated spectral behavior of the source during the mHz oscillations, we analyzed spectra of observations that have GTI segments with tentative QPO detections and compared them with segments in the same observation with non-detections. This selection led to the analysis of the QPO and non-QPO segments from Obs2, Obs4, Obs5, and Obs6. Spectra of the QPO segments were combined for each observation. Similarly all the segments with non-detections were combined for each observation to perform the comparative spectral analysis. We analyzed the 1--10 keV NICER spectra using XSPEC v12.14.1 \citep{arnaud1996}. Spectra below 1 keV were ignored owing to large residuals that are instrumental in origin (e.g., \citealt{sudha2025,li2024}, \citealt{prabhakar2022} and references therein, \citealt{manca2023}). Spectra above 10 keV were ignored due to background dominance. C-statistic was used for spectral fitting. As mentioned in Section~\ref{sec:intro}, the continuum was modeled using a multicolor disk blackbody ({\sc DiskBB}), a single temperature blackbody ({\sc bbody}) and a powerlaw \citep{lin2007,cackett2010}, along with the ISM absorption model ({\sc Tbabs}).

\begin{figure}   
\centering
\includegraphics[width=0.48\textwidth, angle=0]{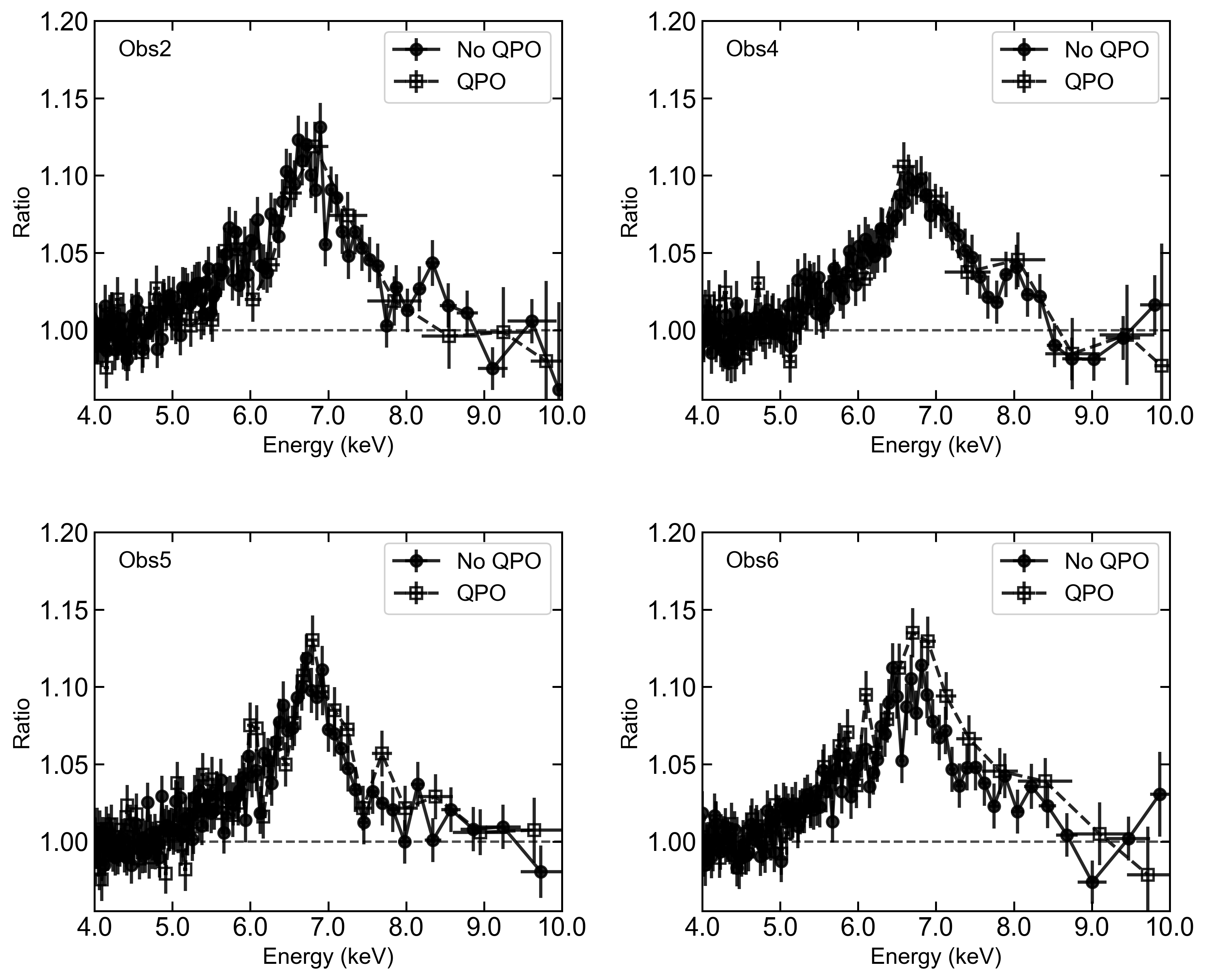}
\caption{Ratio of NICER data to continuum model indicating the presence of a Fe K line emission. The QPO and non-QPO epochs are overplotted for each observation shown in different panels.}
\label{fig:alliron}
\end{figure}

We modeled the QPO epochs and non-QPO epochs simultaneously for each observation. Obs2 had the longest non-QPO epoch and was therefore, used as a reference for the spectral modeling of all the other observations. Continuum modeling revealed absorption edges at $\sim$ 1.84 keV and $\sim$\,1.2\,keV and a broad Fe K$\alpha$ feature at $\sim$ 6.7 keV which is a reflection feature (see Figure \ref{fig:alliron}). The absorption edges are possibly the Silicon K edge\footnote{\href{https://heasarc.gsfc.nasa.gov/docs/nicer/data\_analysis/workshops/2024/joint2024.html}{NICER Data Analysis Workshop 2024}} associated with the NICER detector and Ne K edge associated with the ISM (e.g. \citealt{cackett2008a} and references therein, \citealt{iaria2005}), respectively. We used two {\sc edge} models to account for these features. To account for the reflected emission, we used the {\sc RELXILLNS} reflection model \citep{dauser2014,garcia2014}. For a detailed review of the reflection model along with specific details on the parameters, see \citealt{ludlam2024}. We fixed the reflection fraction to -1 to return only reflected emission and tied the blackbody temperature of the {\sc RELXILLNS} model to the blackbody temperature of the single-temperature blackbody component. To estimate the uncertainties, a Markov Chain Monte Carlo (MCMC) chain was run with 100 walkers, a chain length of 50000 and a burn-in length of 10$^{6}$. Errors reported on the spectral parameters in this work are at the 90\% confidence level. The complete model used for the overall spectrum is {\sc edge*edge*Tbabs(Diskbb+Bbody+RelxillNS+\\ Powerlaw)}. 

Apart from the powerlaw model for the Comptonized emission, the more physical model ({\sc Thcomp}) was also employed to model the Comptonized spectra \citep{zdziarski2020}. The energy range was extended from 0.1~keV to 1000 keV with 1000 logarithmic bins as recommended for this model while fitting. We tried two different models, one with {\sc Thcomp} convolved with {\sc bbody} and the other with {\sc Thcomp} convolved with the {\sc DiskBB}. We attempted these fits using optical depth instead of the photon index parameter in {\sc Thcomp} and fixed its value to 10 (e.g. \citealt{chattopadhyay2024}). We also attempted the fit by allowing the parameter to vary. For all the attempts at modeling using {\sc Thcomp}, the covering fraction and electron temperature tended to unphysical values. Covering fraction tended to either very low values of the order of 10$^{-8}$ or pegged at the highest value of 1. Electron temperature too tended to high values 20--140~keV. Since these model combinations led to unphysical constraints on model parameters, we continued with the simple powerlaw model for the remaining analysis.

Obs2 was used as the reference for modeling the remaining observations as mentioned above. Therefore, for Obs4, Obs5, and Obs6, parameter values of hydrogen column density N$_{H}$, inclination angle, and iron abundance (A$_{Fe}$) were fixed to that obtained from the spectral modeling of Obs2. The emissivity index $q$$_{1}$ was fixed at 3 for all epochs (e.g., \citealt{ludlam2017}) for a flat, Euclidean geometry and a single emissivity profile is used ($q$$_{1}$=$q$$_{2}$). Since the disk density for NS LMXBs is expected to be $>$ 10$^{20}$ cm$^{-3}$ \citep{frank2002,shakura1973}, we fixed the disk density parameter $\log(n_{e})$ at 19 for all epochs, as this is the hard limit of $\log(n_{e})$ in the {\sc RELXILLNS} model.
The edge parameters, A$_{Fe}$, and inclination of the QPO epochs were tied to that in the non-QPO epochs during the simultaneous fitting procedure. See Table \ref{tab:spectral_params} for the best fit results.

\begin{table*}
\centering
\caption{Best-fit spectral parameters of the QPO and non-QPO epochs for each observation where candidate mHz QPOs were detected from the 1--10 keV NICER spectra, fitted using  {\sc edge*edge*tbabs(diskbb + bbody + RelxillNS + Powerlaw)} model. 
The subscript BB represents the {\sc bbody} model and disk represents {\sc diskbb} model. Errors are quoted at a 90\% confidence level.}
\label{tab:spectral_params}
\centering
\scriptsize
\begin{tabular}{llcccccccc}
\hline\hline
Model & Parameter & \multicolumn{2}{c}{Obs2} & \multicolumn{2}{c}{Obs4} & \multicolumn{2}{c}{Obs5} & \multicolumn{2}{c}{Obs6} \\
 &  & No QPO & QPO & No QPO & QPO & No QPO & QPO & No QPO & QPO \\
\hline
edge & $E$ (keV) & \multicolumn{2}{c}{$1.85 \pm 0.01$} & \multicolumn{2}{c}{$1.85 \pm 0.01$} & \multicolumn{2}{c}{$1.86 \pm 0.01$} & \multicolumn{2}{c}{$1.86 \pm 0.01$} \\
     & $\tau$ & \multicolumn{2}{c}{$0.07\pm 0.01$} & \multicolumn{2}{c}{$0.06\pm 0.01$} & \multicolumn{2}{c}{$0.057\pm 0.005$} & \multicolumn{2}{c}{$0.06 \pm 0.01$} \\
edge & $E$ (keV) & \multicolumn{2}{c}{$1.22 \pm 0.02 $} & \multicolumn{2}{c}{$1.14 \pm 0.02$} & \multicolumn{2}{c}{$1.19_{-0.02}^{+0.03}$} & \multicolumn{2}{c}{$1.20_{-0.05}^{+0.03}$} \\
     & $\tau$ & \multicolumn{2}{c}{$0.09 \pm 0.01 $} & \multicolumn{2}{c}{$0.10 \pm 0.02$} & \multicolumn{2}{c}{$0.07 \pm 0.01$} & \multicolumn{2}{c}{$0.08 \pm 0.02 $} \\
\hline
TBabs & $N_{\rm H}$ ($10^{22}$ cm$^{-2}$) & \multicolumn{2}{c}{$1.91_{-0.02}^{+0.03}$} & \multicolumn{2}{c}{--} & \multicolumn{2}{c}{--} & \multicolumn{2}{c}{--} \\
\hline
diskbb & $kT_{\rm in}$ (keV) & $1.78_{-0.07}^{+0.10}$ & $1.79_{-0.12}^{+0.15}$ & $1.79_{-0.10}^{+0.11}$ & $1.74_{-0.14}^{+0.23}$ & $1.74_{-0.12}^{+0.14}$ & $1.78 \pm 0.15$ & $1.76_{-0.11}^{+0.13}$ & $1.72_{-0.11}^{+0.14}$ \\
       & ${\rm norm}_{\rm disk}$ & ${40.6_{-5.0}^{+6.5}}$ & ${44_{-9}^{+10}}$ & ${45_{-7}^{+8}}$ & ${48_{-15}^{+13}}$ & ${46.4_{-9.8}^{+9.4}}$ & ${43.7_{-9.6}^{+12.3}}$ & ${42.2_{-9.3}^{+9.7}}$ & ${43.4_{-9.1}^{+10.8}}$ \\
\hline
relxillNS & q & \multicolumn{8}{c}{3.00 (fixed)} \\
          & $i$ ($^\circ$) & \multicolumn{2}{c}{${30.8_{-3.2}^{+3.1}}$} & \multicolumn{2}{c}{--} & \multicolumn{2}{c}{--} & \multicolumn{2}{c}{--} \\
          & $R_{\rm in}$ ($R_{\rm ISCO}$) & $1.56_{-0.56}^{+0.41}$ & $2.68_{-1.35}^{+1.15}$ & $2.54_{-0.89}^{+0.62}$ & ${5.1_{-9.2}^{+3.2}}$ & $1.78_{-0.55}^{+0.31}$ & ${2.5_{-2.1}^{+0.7}}$ & $1.84_{-0.56}^{+0.39}$ & $2.57_{-1.10}^{+0.97}$ \\
          & $\log(\xi)$ & $2.93_{-0.22}^{+0.17}$ & $2.84_{-0.33}^{+0.39}$ & $2.97_{-0.18}^{+0.14}$ & $2.67_{-0.26}^{+0.36}$ & $2.73_{-0.15}^{+0.21}$ & $2.95_{-0.25}^{+0.21}$ & $2.88_{-0.20}^{+0.21}$ & $2.88_{-0.27}^{+0.23}$ \\
          & $A_{\rm Fe}$ & \multicolumn{2}{c}{${4.8_{-1.7}^{+2.6}}$} & \multicolumn{2}{c}{--} & \multicolumn{2}{c}{--} & \multicolumn{2}{c}{--} \\
          & $\log(n_{e})$ (cm$^{-3}$) & \multicolumn{8}{c}{19 (fixed)} \\
          & ${\rm norm}_{\rm rel}$ & $1.93_{-0.35}^{+0.55}$ & $1.91_{-0.58}^{+0.69}$ & $1.72_{-0.21}^{+0.20}$ & $1.96_{-0.51}^{+0.72}$ & $2.09_{-0.31}^{+0.35}$ & $1.75_{-0.27}^{+0.37}$ & $1.98_{-0.28}^{+0.33}$ & $2.27_{-0.45}^{+0.55}$ \\
\hline
bbody & $kT_{\rm bb}$ (keV) & $2.19 \pm 0.19 $ & $2.15_{-0.22}^{+0.30}$ & $2.14_{-0.10}^{+0.16}$ & $1.97_{-0.13}^{+0.18}$ & $2.06_{-0.12}^{+0.19}$ & $2.03 \pm 0.16$ & $2.03_{-0.12}^{+0.16}$ & $2.00 \pm 0.12$ \\
      & ${\rm norm}_{\rm bb}$ & $0.03 \pm 0.01$ & $0.03 \pm 0.01$ & $0.05 \pm 0.01$ & $0.05_{-0.02}^{+0.01}$ & $0.05 \pm 0.01$ & $0.05 \pm 0.01$ & $0.05 \pm 0.01 $ & $0.04 \pm 0.01$ \\
\hline
powerlaw & $\Gamma$ & $2.27_{-0.19}^{+0.20}$ & $2.46_{-0.30}^{+0.29}$ & $2.51_{-0.18}^{+0.13}$ & $2.54 \pm 0.28$ & $2.29_{-0.17}^{+0.18}$ & $2.34 \pm 0.20$ & $2.20_{-0.16}^{+0.21}$ & $2.17_{-0.17}^{+0.19}$ \\
         & ${\rm norm}_{\rm pl}$ & $1.61_{-0.14}^{+0.17}$ & $1.57_{-0.18}^{+0.22}$ & $1.62 \pm 0.07$ & $1.61_{-0.15}^{+0.10}$ & $1.55_{-0.06}^{+0.07}$ & $1.59_{-0.09}^{+0.12}$ & $1.60_{-0.10}^{+0.09}$ & $1.50_{-0.07}^{+0.08}$ \\
         
\hline
C-stat (d.o.f) && \multicolumn{2}{c}{2184.86 (1773)} & \multicolumn{2}{c}{2075.75 (1776)}  &\multicolumn{2}{c}{1997.44 (1776)} &\multicolumn{2}{c}{2177.87 (1776)} \\
\hline\hline
\end{tabular}
\end{table*}

We obtained an N$_{H}$ value of $\sim$ 1.91 $\times$ 10$^{22}$ cm$^{-2}$ which is in close agreement with previous studies of the source (e.g., \citealt{pintore2015,ludlam2019,neal2023}). The disk temperature kT$_{in}$ is consistent within uncertainties for all the epochs, ranging from $\sim$ 1.7--1.9 keV. We obtained an inclination angle of $\sim$ 31$^{\circ}$, which is in close agreement with values in literature for this source (e.g., \citealt{ludlam2019}). The inner disk radii (R$_{in}$) obtained from the best-fit consistently indicates a higher radius in the QPO epoch when compared to the non-QPO epoch, but the large uncertainties prevent a robust conclusion. 
A$_{Fe}$ is supersolar and estimated to be $\sim$ 4.82. This is consistent with studies that have shown that iron abundance is degenerate with the electron density and therefore, for systems with high disk density, a supersolar iron abundance is commonly noticed (for a detailed discussion see \citealt{tomsick2018,garcia2018}). Blackbody temperature, kT$_{BB}$ ($\sim$ 2 keV) is consistent within uncertainties for all epochs irrespective of them being QPO or non-QPO epochs. The spectra is soft with photon index $\sim$ 2.2--2.5 and is consistent within uncertainties among all epochs.

We estimated the unabsorbed flux in the 3--20~keV energy range for each epoch after extending energies in XSPEC. Using a distance of 6.5 kpc to the source, we estimated luminosities at each epoch. Considering L$_{\rm Edd}$=3.8$\times$ 10$^{38}$~erg~s$^{-1}$ \citep{kuulkers2003}, we estimate that the source has a luminosity of $\sim$~0.14--0.16~L$_{\rm Edd}$ in the 3--20 keV energy band.

\section{Discussion} \label{sec:discussion}

We performed the spectro-temporal study of the bright atoll source GX 3+1 using NICER in the 0.5--10~keV energy band. Temporal analysis revealed the presence of candidate millihertz (mHz) QPOs in the data. We made 8 tentative detections of mHz QPOs with central frequencies ranging from 7--15 mHz. This is the first detection of mHz QPOs in GX 3+1. Spectral analysis was performed for each observation by sectioning the data into QPO and non-QPO epochs to perform comparative analysis. 

\subsection{Millihertz QPOs and their causative mechanism}

So far, mHz QPOs have been detected in 10 NS LMXBs viz.  4U 1636-53 \citep{revnivtsev2001}, 4U 1608-52 \citep{revnivtsev2001}, Aql X-1 \citep{revnivtsev2001}, IGR J17480-2446 \citep{linares2010}, 4U 1323-619 \citep{strohmayer2011}, IGR J00291+5934 \citep{ferrigno2017}, GS 1826-238 \citep{strohmayer2018}, EXO 0748-676 \citep{mancuso2019}, 1RXS J180408.9-342058 \citep{tse2021}, and 4U 1730-22 \citep{mancuso2023} where QPOs were found in the 5-15 mHz frequency range. A large subset of these systems exhibit mHz QPOs with specific and consistent properties. These QPOs were only found in a specific range of X-ray luminosities with L$_{2-20}$ keV  $\sim$ (0.05-3.5) $\times$ 10$^{37}$ erg s$^{-1}$. The detected mHz QPOs also had fractional rms amplitude in the $\sim$ 1-4 \% range. Previous studies have identified a pivot in the rms-energy plot at $\sim$ 3 keV, above which the rms decreases up until 5 keV \citep{revnivtsev2001,altamirano2008a,lyu2020}. For a few of studied sources, mHz QPOs have also exhibited a systematic frequency drift (e.g., \citealt{altamirano2008a,mancuso2019}).
These mHz QPOs are also associated with Type-I X-ray bursts \citep{altamirano2008a} as seen in 4U 1636-53, where it was found that the oscillation frequency decreased over time till they disappeared, leading to the occurrence of a Type-I X-ray burst. \citealt{altamirano2008a} found that for 4U 1636-53 when the QPO frequency was $\geq$ 9 mHz, burst were absent, whereas when the QPO frequency was $\leq$ 9 mHz, bursts would occur.

There are a few outliers among these systems. The 11 Hz pulsar IGR J17480−2446 which is located in the globular cluster Terzan 5 showed $\sim$ 2.8-4.5 mHz QPOs occurring at higher luminosities than other sources $\sim$ L$_{2-50}$ keV  $\sim$ 10$^{38}$ erg s$^{-1}$ \citep{linares2012}. The $\sim$ 8 mHz QPO seen in the accreting millisecond X-ray pulsar IGR J00291+5934 occurred at luminosities of the order of $\sim$ 10$^{36}$ erg s$^{-1}$ in the 0.5--2 keV energy range and was found to be present only in the soft energy band $<$ 2 keV \citep{ferrigno2017}. The interpretation of this observed QPO behavior remains uncertain.

4U 1608-52 showed an anticorrelation between the frequency of the kHz QPOs and the brightness of the mHz oscillations \citep{yu2002}, which is opposite to the positive correlation that is usually seen. This led to the conclusion that the flux variation associated with the mHz QPO arises on the NS surface rather than the accretion flow. 

The proposed mechanism for the origin of mHz QPOs is the marginally stable nuclear burning (MSNB) of Helium on the NS surface \citep{heger2007}. MSNB is an quasi-periodic oscillatory burning mode occurring at the transition between stable and unstable burning regimes. This oscillatory burning of Helium has a period of $\sim$ 2 minutes, which agrees with the range of frequencies detected for mHz QPOs. This oscillation period is approximately equal to the geometric mean of accretion and thermal timescales. According to this theory, the MSNB mode should occur at local accretion rates $\sim$ Eddington rate \citep{bildsten1998}. But observations of mHz QPOs suggest that the onset of stability occurs at accretion rate close to the $\sim$ 0.1 Eddington rate. This contradiction could be explained away if we consider that the accreted material is confined to $\sim$ 10 \% of the NS surface or/and that there is turbulent rotational mixing of the material that transports it to greater column depths \citep{keek2009}. The fact that mHz QPOs occur at specific local accretion rates makes them good tracers of the accretion rate in these systems. 

Both these explanations are observationally supported. Studies have shown that Type-I X-ray bursts occurring after mHz QPOs exhibited lightcurves having positive `convexity' (4U 1636-53; \citealt{lyu2016}). This would happen for bursts occurring at the NS equator \citep{cooper2007, maurer2008}, which suggests that local accretion rate is higher at this point. Based on the studies of 4U 1636-536, \citealt{altamirano2008a} and \citealt{lyu2015} found frequency drift of the mHz QPOs, where the drift timescale was suggested to be  associated with cooling of the deeper layers. Frequency drifts were also seen in EXO 0748–676 \citep{mancuso2019}, 4U 1608–52, and Aql X–1\citep{mancuso2021}.

\subsection{mHz QPOs in GX 3+1: A comparative study}

We performed a comparison of the basic mHz QPO properties as seen in literature (see Table 1 in \citealt{tse2021}) with our results. We note that these candidate QPOs were detected in the soft spectral state as inferred from the spectral fit (i.e., continuum described by two thermal components and a soft powerlaw) and the HID. This is well in agreement with the spectral states determined for other sources during mHz QPO detections (e.g., \citealt{revnivtsev2001, altamirano2008a, strohmayer2018, lyu2019, mancuso2023}). Based on the results from the spectral analysis we note that the luminosity (3--20 keV) at which we tentatively detect the mHz QPOs range from 5.3-6.1$\times$ 10$^{37}$~erg~s$^{-1}$. This corresponds to $\sim$~0.14--0.16~L$_{\rm Edd}$. And our timing analysis shows a fractional rms range of 0.48--1.41\% for the mHz frequency range of $\sim$ 7--15 mHz. 
\begin{figure*}[!ht]   
\centering
\includegraphics[width=0.95\textwidth, angle=0]{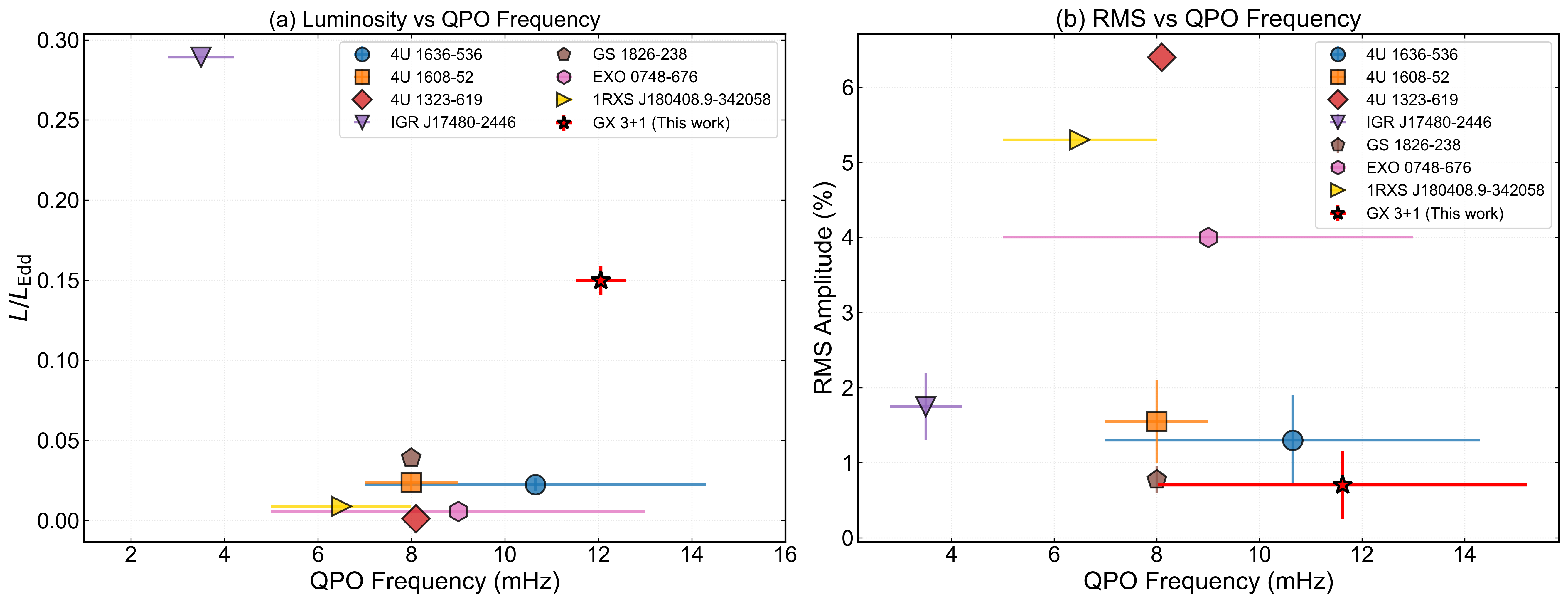}
\caption{Left panel shows the distribution in luminosities of tentative detections in Eddington units with respect to frequency. Right panel shows the spread in fractional rms amplitudes of candidate mHz QPOs with respect to frequency. Results for GX 3+1 (this work) is marked with a `*' symbol. Values for other sources are taken from \citealt{revnivtsev2001, altamirano2008a, strohmayer2011, linares2012, strohmayer2018,mancuso2019,tse2021}}
\label{fig:comparison}
\end{figure*}
We performed an energy dependent study of the fractional rms amplitude of the candidate mHz QPOs in GX 3+1. Most studies indicate an increase in fractional rms amplitude of mHz QPOs with energy till $\sim$ 3 keV (e.g., \citealt{strohmayer2018, lyu2020, mancuso2023}). Therefore, \cite{lyu2020} suggested that the increase in rms with energy till $\sim$ 3 keV could be a characteristic feature of mHz QPOs. 

Studies have also shown that rms amplitude decreases with energy above 3 keV or remained constant within uncertainties \citep{revnivtsev2001, lyu2020}. For certain sources studied using RXTE, such as 4U~1608-52, 4U~1323-619, Aql X-1, and EXO 0748-676, these trends in rms could not be confirmed owing to the energy range in which RXTE operated ($>$ 2~keV). A recent study by \cite{xiao2025} showed an increase in rms amplitude with energy even above 3 keV for GS~1826-238 which was studied using NICER.

As can be seen in Figure \ref{fig:rmsvsenergy} and mentioned in Section \ref{sec:results}, we notice a steady increase in rms amplitude with energy for the 7.8 mHz, 10 mHz, and 11.5 mHz QPOs in GX 3+1. A similar trend is seen in the rms amplitude for the 10.9 mHz QPO till $\sim$ 5 keV and for the 11.4~mHz QPO till 3--4 keV. For both these candidate QPOs the rms amplitude is constant within uncertainties above these pivot points in energy. Meanwhile for the $\sim$ 15 and 14~mHz QPOs, the uncertainties are too large to notice or confirm a trend.

Based on the model by \cite{heger2007}, a variation in accretion rate could lead to changes in the amplitude of mHz oscillations. An increase in mass accretion rate could lead to a stable burning regime, which would result in the lowering of QPO amplitudes and higher QPO frequencies \citep{xiao2025}. But \cite{lyu2019} found no association between rms amplitude and the S$_{a}$ parameter (HID track parameter) in 4U 1636-53, which increases with accretion rate \citep{hasinger1989}. \citealt{xiao2025} found that 
GS 1826-238 was found to be in a soft state during the detection of mHz QPOs and found no association between the QPO presence and the position of the source in the HID. In contrast, for other sources such as 4U 1730−22 \citep{mancuso2023} and EXO~0748−676 \citep{mancuso2019}, mHz QPO detections occurred when these source were in the low hard state or a high count rate state. However, they conclude that this discrepancy could be due to the limited luminosity range (0.10-0.13 L$_{Edd}$) and minimal associated spectral evolution of GS~1826-238. Our study also shows no association between rms amplitudes and location of the tentative mHz detections on the HID, where the location on the HID could be considered as a proxy for mass accretion rate. But this could be a consequence of a limited sampling of the HID track. An alternative scenario would be that this lack of correlation could be attributed to the fact that S$_{a}$ traces the global accretion rate whereas the mHz amplitude would depend on the local mass accretion rate.

Simulations by \cite{keek2009} showed that the rms amplitude of mHz QPOs increased with a decrease in the heat flux from the NS crust. Their study also showed that upon considering the turbulent chemical mixing of the accreted material, the amplitude of the oscillatory burning mode could be higher. But currently, both these models do not address the energy dependence of the rms amplitude of these oscillations. 

\cite{lyu2020} suggests that, since the color temperature of Type-I X-ray bursts is around 2 keV and the pivot point seen in the rms-energy relation of mHz QPOs seen in their study is also around $\sim$ 2--3 keV, this could possibly indicate that these two phenomena share a common origin mechanism on the NS surface. But our results indicate a higher pivot point in energy in the energy-rms relation, which challenges this idea. Also, the lack of Type-I X-ray bursts in the data prevents any further exploration of this possibility. Type-I X-ray bursts were not observed for this source in the study by \cite{gnarini2024}. Future observations from facilities such as eXTP \citep{zhang2025b} could help further exploration of this aspect.

It is possible that additional mechanisms or contribution from non-thermal components could play a role here. But in its current state, the existing models cannot fully explain the energy-rms relation for mHz QPOs and this suggests that the models should be further expanded to address the observational results.

The frequency range in which we have tentatively detected mHz QPOs and the fractional rms amplitude of these detections are consistent with what is noted in literature as shown in Figure \ref{fig:comparison}.

But we note that the luminosity at which we have tentatively detected mHz QPOs ($\sim$ 0.14--0.16 L$_{\rm Edd}$) is higher than that for other sources in literature (L $\sim$ 0.1--0.13 L$_{\rm Edd}$) with an exception of IGR~J17480-2446. This source exhibited mHz QPOs at higher luminosities of L $\sim$ 0.4--0.5~L$_{\rm Edd}$ \citep{linares2012}. But to enable a direct comparison between the luminosity estimated during our tentative detections and the luminosities reported for mHz QPOs in earlier studies, we must ensure that they are estimated in the same energy band. 

We consider 3--20 keV to be the common energy band for flux comparison. For sources that have spectral parameter values available in literature, we replicated their spectral models within XSPEC and estimated flux in the 3--20 keV energy band. This analysis was performed for IGR J17480-2446 based on \cite{linares2012}, 1RXS~J180408.9-342058 based on \cite{ludlam2016}, and GS 1826-238 based on \cite{strohmayer2018} and \cite{xiao2025}. For 4U 1323-619 and EXO 0748-676, we used archival RXTE observations mentioned in   \cite{strohmayer2011} and \cite{mancuso2019} (their observation ``B") and adopted models from \cite{gambino2016} and \cite{sidoli2005} for modeling the spectra. For this we used the `pcaprepobsid' and `pcaextspect2' tools that are made available to analyze RXTE PCA Standard2 data and extract the spectra respectively. Flux was estimated in the 3--20 keV energy band after obtaining the best-fit. Figure~\ref{fig:comparison} shows these estimated luminosities in the left panel and it shows fractional rms amplitudes of mHz QPOs detections in different sources from previous studies on the right panel.

As per \cite{heger2007}, MSNB is expected to occur close to the Eddington accretion rate. We could say that the higher luminosity in our tentative detections is closer to what would be expected in the case of the MSNB mechanism compared to the other sources, although it is still significantly lower than expected. To explain the occurrence of mHz QPOs at higher luminosities in IGR J17480-2446, \cite{linares2012} suggested the possibility that these mHz QPOs could be occurring at the boundary of stable Helium burning, whereas for the low-luminosity mHz QPOs could be occurring at the boundary of stable Hydrogen burning. We could consider this possibility for our tentative detections as well, although we cannot expand or test this possibility.

But we must caution that the luminosity estimate for GX 3+1 carries  uncertainty owing to the source distance not being precisely constrained and the spectral fits being estimated over a limited energy passband. Considering these caveats, we must refrain from drawing strong conclusions about the behavior of GX 3+1 and instead suggest that the apparent discrepancy warrants further investigation.

Spectral analysis of the QPO and non-QPO epochs indicate parameter values of thermal components and reflection parameters to be consistent between both the epochs within error bars. Therefore, we do not find any robust trends in parameter variation. Reflection modeling indicates a disk close to the NS with R$_{in}$ extending down to ISCO upon considering uncertainties. This is in agreement with R$_{in}$ values from literature. But we must note that since the reflection spectral model is constrained entirely in the limited NICER bandpass below 10 keV, the constraints on the reflection spectral model parameters that are sensitive to the high-energy spectral shape, in particular the iron abundance, ionization parameter, and inner disc radius are possibly not fully quantified within the available bandpass, and should therefore be interpreted with caution.

Based on the highest blackbody normalization parameter obtained from a QPO epoch (Obs5), we estimate the emitting blackbody radius by assuming a spherical emission region. We obtain an emission radius estimate of 8.97 $\pm$ 1.67 km which is close to the canonical NS radius. A color-correction factor of 1.7 and distance of 6.5~kpc was used for this estimation. This estimate would be for the entire NS surface rather than the localized burning region alone. In this case, without an estimate of the radial extent of the burning zone itself, it would be difficult to reconcile with the \cite{heger2007} model and resolve the question of the mHz QPOs' occurrence in view of the local vs global mass accretion rates.

\section{Conclusion} \label{sec:conclusion}

We discovered millihertz QPOs in the bright atoll source GX 3+1 using NICER in the 0.5--10 keV energy band. We obtained a fractional rms amplitude range of 0.51-1.41 \% for the mHz frequency range of $\sim$ 7-15~mHz. The mHz QPO properties seen in our study are well in agreement with previous detections of such QPOs in NS LMXBs, except for the luminosity at which these mHz QPOs occur (relying on the current distance estimate). Our study reveals that the tentative detections of mHz QPOs in GX 3+1 occur at luminosities that is twice the range of luminosities shown by other such detections in NS LMXBs. 
We find no association between the HID location of candidate mHz QPOs and their rms amplitudes. The rms-energy relation of the tentatively detected mHz QPOs and the luminosity at which they occur warrants further investigation into the behavior of mHz QPOs in GX 3+1. Here we note that the observational sample of NS LMXB sources exhibiting mHz QPOs remain relatively small and therefore a fully self-consistent framework is yet to be established. The results presented here for GX 3+1 should therefore be viewed as an addition to this still-evolving picture and as more sources are studied in detail, a more complete and self-consistent picture of the mHz QPO phenomenon is expected to emerge. This source is now added to the existing short list of NS LMXBs exhibiting mHz QPOs.


\section{Acknowledgements}
We thank the anonymous referee for their crucial feedback
that has enhanced the scientific rigor of the manuscript.
This research is supported by NASA under grant 80NSSC25K0096. M.N. is a Fonds de Recherche du Quebec – Nature et Technologies (FRQNT) postdoctoral fellow. This research has made use of data and/or software provided by the High Energy Astrophysics Science Archive Research Center (HEASARC), which is a service of the Astrophysics Science Division at NASA/GSFC.  

\end{document}